# Thermal conductivity of graphene mediated by strain and size

Youdi Kuang[1], Lucas Lindsay[2], Sanqiang Shi[3], Xinjiang Wang[1], Ruiqiang Guo[1],

Baoling Huang[1*]

[1]*Department of Mechanical and Aerospace Engineering, The Hong Kong University of Science and Technology, Clear Water Bay, Kowloon, Hong Kong*

[2]*Materials Science and Technology Division, Oak Ridge National Laboratory, Oak Ridge, Tennessee 37831, USA*

[3]*Department of Mechanical Engineering, the Hong Kong Polytechnic University, Hung Hom, Kowloon, Hong Kong*

## Abstract

Based on first-principles calculations and full iterative solution of the linearized Boltzmann-Peierls transport equation for phonons within three-phonon scattering framework, we characterize the lattice thermal conductivities $k$ of strained and unstrained graphene. We find $k$ converges to 5450 W/m-K for infinite unstrained graphene, while $k$ diverges for strained graphene with increasing system size at room temperature. The different $k$ behaviors for these systems are further validated mathematically through phonon lifetime analysis. Flexural acoustic phonons are the dominant heat carriers in both unstrained and strained graphene within the temperature considered. Ultralong mean free paths of flexural phonons contribute to finite size effects on $k$ for samples as large as 8 cm at room temperature. The calculated size-dependent and temperature-dependent $k$ for finite samples agree well with experimental data, demonstrating the ability of the present approach to predict $k$ of larger graphene sample. Tensile strain hardens the flexural modes and increases their lifetimes, causing interesting dependence of $k$ on sample size and strain due to the competition between boundary scattering and intrinsic phonon-phonon scattering. These findings shed light on the nature of thermal transport in two-dimensional materials and may guide predicting and engineering $k$ of graphene by varying strain and size.

* Corresponding author. Tel.: Fax: +852-23587181. E-mail address: mebhuang@ust.hk (Baoling Huang).

**Introduction**

Graphene, a two-dimensional (2D) sheet of carbon atoms, has attracted great interest in recent years due to its extraordinary mechanical, chemical, electronic and thermal properties. The experimentally reported ultrahigh thermal conductivity $k$ (up to 5300 W/m-K [1]) of graphene renders it the most thermally conductive material, and of promise for thermal management applications [2-4]. Moreover, graphene provides a benchmark model for the study of thermal transport in 2D materials. Therefore, intensive efforts have been committed to understand the underlying thermal transport physics in graphene experimentally [1,5-16] and theoretically [2,17-30].

Previous experimental studies [1,5,6,10] using an opto-thermal Raman technique have reported $k$ values of suspended graphene that scatter significantly, ranging from 600 [10] to 5300 [1] W/m-K even with temperature effects accounted for. Using direct thermal-bridge measurements, Xu *et al.* [7] recently observed significant size effects on $k$ of graphene and reported that $k$ diverges logarithmically with sample length; in contrast, previous Raman experiments did not discern a size-dependence of $k$ [1,6]. Pettes *et al.* [11] found that $k$ of graphene may be significantly influenced by the residual polymeric layer produced during the transfer process of graphene. It can be seen from previous experiments that measured $k$ depends strongly on various extrinsic factors including sample size, process conditions, sample quality, measurement method and substrate coupling [12,13], all giving large scatter in experimental $k$ data.

On the theoretical side, fundamental problems concerning the details of thermal transport in graphene have been subjects of debate [17-30], including the convergence behavior of $k$ with system size, the extent of the diffusive and ballistic transport regimes, the role of flexural acoustic (ZA) phonons for thermal transport and strain effects on the convergence of $k$. It is generally believed that acoustic phonons [31] dominate the thermal transport in graphene. Based on this, 2D models give a logarithmic divergence with system size [27] but neglect the contributions from ZA phonons due to their low group velocities near the center of first Brillouin zone (FBZ) and their large Grüneisen parameters [32]. However, molecular dynamics (MD) simulations [17,18,26,29,30] with large system sizes have reported that room-temperature $k$ of graphene converges, though at values much lower than experimental results on finite-size systems. Usually, the accuracy of classical MD simulations heavily relies on the quality of the interatomic potentials used, e.g., qualitatively opposite MD predictions on the convergence of $k$ for strained graphene were reported by two groups [18,30] using different empirical potentials.

First-principles lattice dynamics calculations [22-24,28] of $k$ of graphene within three-phonon scattering framework have also been conducted. Using the single-mode relaxation time approximation (SMRTA), Bonini *et al.* [22] showed that $k$ of infinite graphene diverges under infinitesimal isotropic tensile strains, while $k$ converges to ~550 W/m-K for infinite unstrained graphene at room temperature. The results are partly different with aforementioned MD predictions [18] that *k* of infinite graphene diverges only under large tensile strain (> 0.02). We note that the SMRTA incorrectly treats the momentum-conserving Normal (N) processes as independent resistive processes on the same footing as Umklapp (U) processes [33], and it can not be used to appropriately present the phonon thermal transport in graphene, as justified by Lindsay *et al.* [23] and Fugallo *et al.* [28]. Based on a full iterative solution of the linearized Boltzmann-Peierls Equation (BPE), Lindsay *et al.* [23] found that ZA phonons give the dominant contributions to *k* in finite graphene up to 50 μm with strong dependence of *k* on boundary scattering. They also showed that *k* is relatively insensitive to small tensile strain. Using a similar approach, Fugallo *et al.* [28] argued that tensile strain will not cause a divergence of *k*, and predicted qualitatively different dependences of *k* on strain from the full BPE solution and the SMRTA; however, the corresponding physical explanations are lacking.

In this work, we intend to elucidate these discrepancies using a rigorous first principles BPE for phonon transport approach. Full iterative solutions of the linearized BPE from reciprocal-space calculations and further mathematical analysis show that with increasing system size *k* converges for unstrained graphene and diverges for strained graphene. Mode contribution analysis shows that ZA phonons are the major heat carriers and control the convergence behaviors in both unstrained and strained graphene up to 3000 K. Further, the long mean free paths of ZA phonons make finite size effects on *k* persistent up ~8 cm for unstrained graphene. The joint effect of strain and size on *k* of finite graphene is also discussed.

A microscopic description of the lattice thermal conductivity *k* can be derived from BPE for phonons [21,22,24,34] in three-phonon scattering framework. Considering the isotropic thermal conductivity of infinite graphene, along an in-plane crystallographic direction $\alpha$, the intrinsic *k* is calculated by $k = k^{\alpha\alpha} = \frac{1}{k_B T^2 V N_1^2} \sum_\lambda f_\lambda (f_\lambda + 1)(\hbar\omega_\lambda)^2 v_\lambda^\alpha v_\lambda^\alpha \tau_\lambda$ [23,34], where $\omega_\lambda$, $v_\lambda^\alpha$, and $\tau_\lambda$ are the angular frequency, group velocity and phonon lifetime, respectively. Here $\lambda$ represents a phonon mode with wavevector and branch index. $k_B$, $\hbar$, $f_\lambda$ are the Boltzmann constant, the reduced Plank constant and the Bose-Einstein distribution of phonons at temperature $T$, respectively. $V$ is the volume of the graphene unit cell with a thickness of 0.335 nm [23]. This work combines an iteratively self-consistent solution to the linearized BPE with harmonic and anharmonic interatomic force constants (IFCs) from Density Functional Perturbation Theory (DFPT) and Density Functional Theory calculations [23,24], respectively, using the QUANTUM ESPRESSO package [35] within the local density approximation and using a norm-conserving pseudopotential to represent the core carbon electrons. To determine the harmonic IFCs, DFPT calculations are employed with a 13×13 $k$-point mesh and 120 Ryd plane-wave cutoff for the 2-atom unit cell. To determine the interatomic forces and resulting anharmonic IFCs, DFT calculations with Γ-point sampling in slightly perturbed 162-atom supercells with a 100 Ryd plane-wave cutoff are used. Interactions are considered out to fifth nearest neighbors of the unit cell atoms and crystal symmetries, translational and other invariance conditions are enforced on the IFCs. Further technical details for the calculations of dispersion relations and three-phonon scattering rates can be found in [23,34]. Comparison of the calculated dispersion of graphene with experimental data [36,37] gives excellent agreements. The full 2D FBZ (Fig.1(a)) is discretized into a Γ-centered regular $N_1 \times N_1$ grid with $N_1$ up to 501 considered in this study.

Fig. 1(a) shows the calculated room-temperature $k$ of unstrained graphene (isotropic tensile strain $\varepsilon = 0$) with respect to the $q$-point sampling density. Here, $\varepsilon=(a-a_0)/a_0$ where $a_0=2.44$Å is the calculated equilibrium lattice constant and $a$ is the lattice constant for a given tensile strain. For each $N_1$, an iteration precision of $1\times10^{-5}$ (difference of $k$ values for successive iterative steps) is taken to ensure full self-consistent convergence of $k$. Interestingly, $k$ decreases with increasing number of modes and a grid-converged $k = 5450$ W/m-K is achieved for $N_1 \geq 301$. This $k$ value from the iterative approach is several-fold higher than those from the SMRTA [22,24], confirming previous findings [21,23] that both N- and U-processes and their relationship influencing the nonequilibrium populations of phonon modes are important for determining $k$ of graphene. The convergence of $k$ in our calculations for infinite unstrained graphene also justifies that intrinsic three-phonon scatterings can confine $k$, i.e., higher-order inter-phonon scatterings are not required for convergence of $k$ as was previously suggested for $k$ of unstrained single-walled carbon nanotubes [38,39]. Under different strains $\varepsilon = 0.0025$, 0.01 and 0.1, $k$ increases nearly linearly with increasing $N_1$ and at a fixed $N_1$ a larger $\varepsilon$ gives a higher $k$, indicating non-negligible contributions from longer-wavelength phonon modes and the divergence of $k$ with system size under strain. This is consistent with the SMRTA-based prediction by Bonini *et al*. [22]. Unlike the work of Ref. 22, anharmonic IFCs are calculated here for each strain value considered. With increasing strain the magnitudes of the anharmonic IFCs tend to decrease. Our calculations show that neglecting strain effects on these IFCs does not change the convergence behavior qualitatively but causes significant underestimation of $k$ especially for large strain. For example, a 31-fold underestimation in $k$ at $\varepsilon = 0.1$ for $N_1= 301$ is observed when using unstrained anharmonic IFCs. Moreover, unlike the work of Ref. 28, we find qualitatively uniform dependences of $k$ on strain from the full BPE solution and the SMRTA, consistent with that found in strained 3D diamond [40]. Unless specified otherwise, all the results shown below are for $N_1=301$ for unstrained and strained graphene.

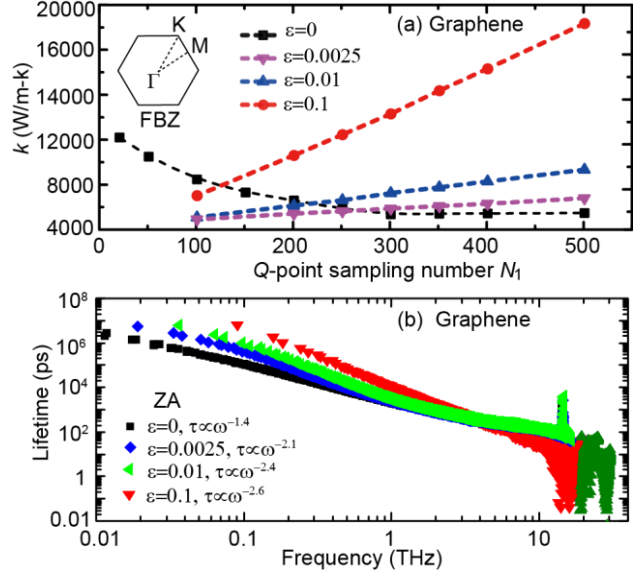

Fig. 1 (a) Convergence of $k$ of graphene with $q$-point sampling density for different isotropic tensile strains. (b) Lifetimes of ZA phonons in graphene under different strains.

Fig.1(b) shows the calculated transport lifetimes $\tau_\lambda$ of ZA acoustic phonons for strained and unstrained infinite graphene at room temperature. Here we only consider ZA phonons since they dominate thermal transport in graphene. For unstrained graphene, $\tau_\lambda \propto \omega_\lambda^{-1.4}$ is found for low-frequency ZA phonons. In strained graphene, ZA phonon lifetimes are more strongly dependent on frequency, having $\tau_\lambda \propto \omega_\lambda^{-2.1}$, $\omega_\lambda^{-2.4}$ and $\omega_\lambda^{-2.6}$ for $\varepsilon = 0.0025$, 0.01 and 0.1, respectively. The lifetime enhancement is due partly to decreasing magnitude of the anharmonic IFCs with tensile strain [23], which reduces the scattering matrix elements, and partly to the reduction in ZA phonon density of states caused by zone center dispersion linearization [22], which leads to less scatterings of ZA phonons [18]. These lifetime data provide a better understanding of the results shown in Fig. 1(a). The nonzero minimum $|q|$ corresponding to $N_1$ is $|q| = \frac{\gamma}{N_1}$ where $\gamma$ is a constant. Then, at small wavevector the phonon frequency and velocity corresponding to $|q|$ in the ZA branch [22] are given by $\omega_\lambda^2 = C|q|^2 + D|q|^4 = \frac{C\gamma^2}{N_1^2} + \frac{D\gamma^4}{N_1^4}$ and $v_\lambda^\alpha = \frac{d\omega_\lambda}{dq}$. The lifetimes follow a power law behavior $\tau_\lambda = A\omega_\lambda^{-\beta}$. Here $A$, $C$, $D$ and $\beta$ are strain-dependent parameters and $C = 0$ for zero strain. The contribution of mode $\lambda$ to $k$ is $k_\lambda = \frac{f_\lambda(f_\lambda+1)(\hbar\omega_\lambda v_\lambda^\alpha)^2 \tau_\lambda}{k_B T^2 V N_1^2}$. For unstrained and strained graphene as $N_1 \to \infty$, $\omega_\lambda \to 0$ and $f_\lambda \to \frac{k_B T}{\hbar\omega_\lambda}$. Therefore, for ZA mode $\lambda$ in unstrained graphene ($\beta = 1.4$), we have $\lim_{\substack{\omega_\lambda \to 0 \\ N_1 \to \infty}} k_\lambda \propto \lim_{N_1 \to \infty} \frac{1}{N_1^{4-2\beta}} = 0$, implying ZA phonons in the long-wavelength limit contribute little to $k$ and $k$ will converge. For strained graphene ($\beta > 2$), we have $\lim_{\substack{\omega_\lambda \to 0 \\ N_1 \to \infty}} k_\lambda \propto \lim_{N_1 \to \infty} N_1^{\beta-2} \to \infty$, indicating $k$ will diverge. Bonini *et al.* [22] showed that frequency-dependent $\tau_\lambda^0$, SMRTA phonon lifetimes, of the ZA phonons satisfy the power law $\tau_\lambda^0 \propto \omega_\lambda^{-\beta}$ in the long-wavelength limit, with $\beta = 1$ from an analytical approach and $\beta > 2$ from numerical results for unstrained and strained ($\varepsilon$ = 0.005) graphene, respectively. The SMRTA underestimates $\beta$ due to the dominance of N-processes. The fitted $\beta$ parameters in our calculations support the corresponding conclusions regarding the convergence of unstrained and strained graphene. Further, our calculations with larger $N_1$ ($N_1$=401, 501) gives slightly larger $\beta$ values at low frequencies than that corresponding to $N_1$=301 for graphene with strain $\varepsilon$ = 0.01. The reason is as follows: larger $N_1$ involves lower frequencies for which the dispersion relations have stronger linearization than those of high-frequency phonons. Therefore, the corresponding $\beta$ value based on these lower frequencies is larger, and still gives divergent behavior. This further validates our conclusion that $k$ of strained graphene diverges.

We also compare the calculated $k$ of unstrained graphene with measured data for graphene of varying sample size, temperature and isotope abundance. Here we discuss the implementation of extrinsic scattering mechanisms in the calculation of $k$ for better comparison with measured $k$ values. We define $k$ as a scalar value, ignoring anisotropy from finite system size due to the relatively large experimental samples that we are comparing with [23]. The lifetime $\tau_\lambda^f$ of a phonon in a finite sample may be calculated using the Matthiessen rule [23], expressed here as $\frac{1}{\tau_\lambda^f} = \frac{1}{\tau_\lambda} + \frac{1}{\tau_\lambda^{iso}} + \frac{1}{\tau_\lambda^b} + \frac{1}{\tau_\lambda^w}$. $\frac{1}{\tau_\lambda}$ is the intrinsic phonon-phonon scattering rate; $\frac{1}{\tau_\lambda^{iso}}$ represents the scattering rate from naturally occurring isotopes (1.1% $C^{13}$) in graphene and is obtained from perturbation theory for a random isotope distribution [34]; $\frac{1}{\tau_\lambda^b}$ represents the scatting rate by contact boundaries and is expressed empirically as $\frac{1}{\tau_\lambda^b} = \frac{2|v_\lambda^x|}{L}$ [38] and $\frac{1}{\tau_\lambda^b} = \frac{|v_\lambda^x|}{D}$ [23,28] respectively for a rectangle sample and a circular sample. This is consistent with experiments using rectangular samples in the thermal-bridge measurements of $k$ [7] and circular samples in the Raman measurements of $k$ [6]. The direction of the temperature gradient, $x$, is assumed to be along the sample length, $L$, or diameter, $D$. $\frac{1}{\tau_\lambda^w}$ represents the scatting rate due to finite sample width $W$ of the rectangular samples and is expressed as [41] $\frac{1}{\tau_\lambda^w} = \frac{2|v_\lambda^y|}{W}$, where $v_\lambda^y$ is the group velocity along the width direction and perpendicular to the transport direction. As shown in Fig. 2(a), the calculated room temperature $k$ for different sample lengths $L$ agree well with recently measured data for suspended samples with a width $W$=1.5 μm [7]. Fig. 2(b) compares the calculated $k$ with experimental data from Ruoff's group [6] for graphene suspended over circular wells with a diameter $D$=2.8 μm at different temperatures. For samples with naturally occurring C isotopes, good agreement is observed throughout the considered temperature range. For isotopically purified samples, a ~12% enhancement in $k$ is predicted at room temperature, significantly less than the measured enhancement. Nonetheless, the calculated enhancement in $k$ falls within the experimental uncertainties [6] from the Raman technique. Recently, slight $k$ enhancements with isotopic purification, ~13% and ~16%, were also reported by Lindsay *et al.* [23] and Fugallo *et al.* [28], respectively.

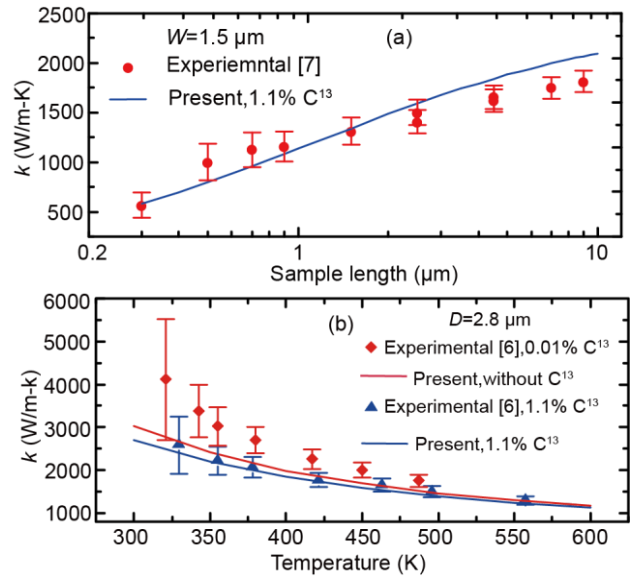

Fig. 2 (a) Comparison of calculated $k$ with previous experimental data for samples of width $W$=1.5 μm and different lengths. (b) Comparison of calculated $k$ with previous experimental data for samples of a fixed diameter $D$=2.8 μm at different temperatures.

To better understand the role of intrinsic three-phonon scattering, we calculate the frequency $\omega_\lambda$ dependent normalized $k$ accumulation for unstrained infinite graphene at 300 K. The values are normalized by the $k$ of unstrained, isotopically pure, infinite graphene, 5450 W/m-K. As shown in Fig.3(a), ZA phonons are the main heat carriers in suspended unstrained graphene, contributing about 88% to $k$ at room temperature. Moreover, the relative contribution of ZA phonons to $k$, as denoted by $k_{ZA}/k$ at given $T$ is shown in Fig 3(b). Although increasing $T$ decreases $k_{ZA}/k$, the dominant contribution of ZA phonons is still obvious (above 80%) for temperatures from 200 K to 3000 K. We also see that the intrinsic $k$ shows different temperature dependent behaviors: $k \propto T^{-1.13} + 6 \times 10^6 T^{-3.96}$ for $T \leq 1200$ K and $k \propto T^{-1.05}$ for $T > 1200$ K. The trends imply that $k$ diverges at 0 K and the contribution of ZA phonons dominate approaching 0 K. On the other hand, the absorption process ZA+ZA→ TA (LA) [22] dominates the scattering of ZA modes and results in decreasing and increasing $k$ contributions of ZA and TA(LA) modes, respectively, with $N_1$ until convergence. For strained graphene, the results show that ZA phonons also provide the dominant contributions to $k$ over the considered temperature range and $k$ still diverges at high temperature up to 3000 K, as seen in Fig.3(c), demonstrating that high temperature does not confine the intrinsic $k$ to converge though the scattering rates increase significantly. As explained by Lindsay *et al.* [19,23], this dominance of the ZA phonons arises because the reflection symmetry of graphene forbids three-phonon scatterings involving odd numbers of ZA phonons, restricting the phase space for ZA phonon scattering. Previous experiments on supported graphene [12,13] also attribute measured reductions in $k$ to the suppression of ZA phonon contributions by substrate coupling. Therefore, engineering $k$ by suppressing the ZA phonon contributions through substrate coupling [12,13], irregular doping or defect distributions [11,28], or irregular out-of-plane deformations may be worthwhile. We note the temperature effect on the *k* behavior of strained graphene is an open question, as mentioned previously by Bonini *et al*. [22]. The most recently work [42] only investigated the temperature effects on phonon thermal transport in unstrained graphene under 800 K, which is still within the low-temperature range considering the graphene Debye temperature is up to 2200 K. Therefore, the present investigations may further our understanding of temperature effects.

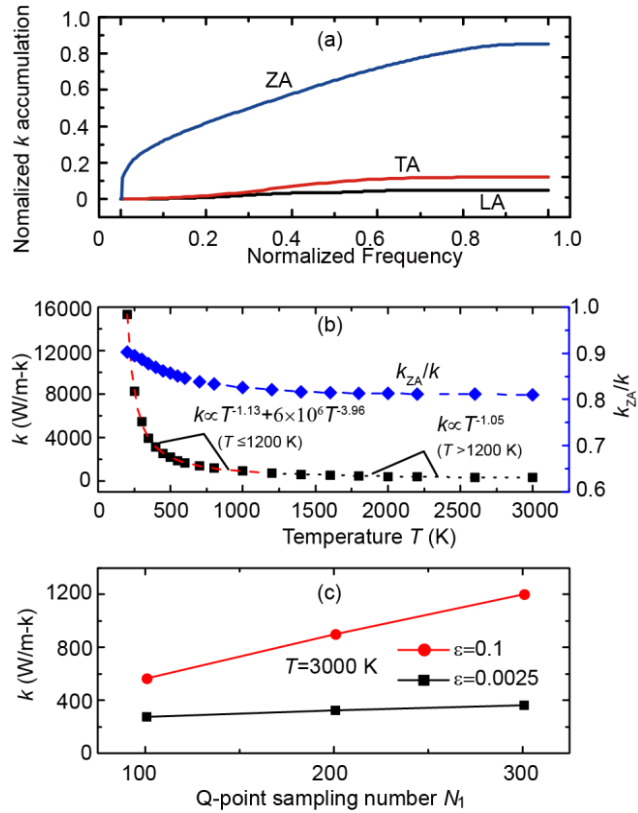

Fig. 3 (a) Normalized *k* accumulations of acoustic phonons for unstrained graphene with respect to frequency normalized by the corresponding cutoff frequency, i.e., 16.1 THz, 23.5 THz and 40.4 THz for ZA, TA and LA branches, respectively. (b) Effects of temperature on the intrinsic *k* and the contribution of ZA phonons. (c) Convergence of *k* with *q*-point sampling density for isotropic tensile strains $\varepsilon=0.0025$ and $\varepsilon=0.1$ at temperature 3000 K.

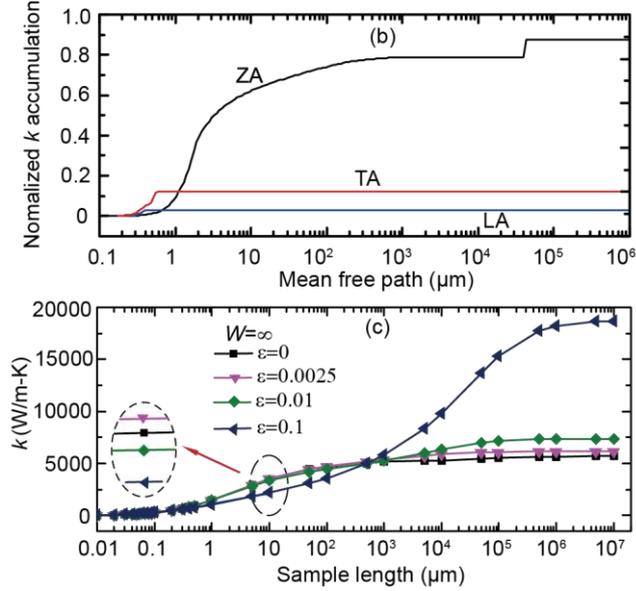

Fig. 4 (a) Normalized $k$ accumulation with respect to the phonon mean free path. (b) Effects of sample length and strain on $k$.

The ratios of phonon transport lifetimes to relaxation times given by SMRTA solutions show significant amplification for most acoustic modes, demonstrating that the SMRTA severely overestimates the intrinsic resistance in graphene. Therefore, using the SMRTA to determine size-dependent $k$ may be misleading as the boundary scattering is relatively weak in comparison [43]. To examine further the mechanism for the significant size effects observed in previous experiments [7] and numerical calculations [23,28,29], we plot the normalized $k$ accumulation with respect to the mean free path (MFP) of acoustic phonons in infinite unstrained graphene without isotope scattering (Fig. 4(a)). The contributions from ZA phonons with ultra-long MFP saturate at about 8 centimeters, while those of in-plane TA and LA phonons saturate around 10 μm or less. Therefore, for suspended samples with the system sizes of several hundreds of microns or less, boundary scattering will significantly limit the thermal conductivity, and the calculated $k$ will increase significantly with increasing system size as demonstrated by the length dependent $k$ shown in Fig. 4(b) corresponding to the case $\varepsilon=0, W=\infty$. We note that the jump in $k$ at ~4 cm in Fig. 4(a) is due to the discrete $q$-sampling mesh, as also seen in previous lattice dynamics calculations [44]. For two neighbor low-frequency ZA modes, their group velocities differ slightly while their lifetimes can differ significantly as shown in Fig.1(b). This results in contributions to $k$ that can differ significantly and give abrupt contributions to the $k$ accumulation in MFP plots. The amplitude (~10%) of the jump (actually the accumulation of a series of small jumps) is consistent with the drastic increase of ZA contribution for frequencies below 0.5 THz as shown in Fig. 3(a).

To illuminate tensile strain effects on $k$ of finite graphene samples, i.e., the interplay between strain and size effects, we also plot the sample length $L$ dependent room-temperature $k$ at different strain levels in Fig. 4(b). Within the applicable length range 1 μm < $L$ ≤ 500 μm, a rather small strain ($\varepsilon=0.0025$) leads to a slightly higher $k$ compared with that of the unstrained case, while further increasing strain results in decreasing $k$, as shown in the inset. Lindsay *et al*. [23] also observed this behavior for a 10-μm system. However, within this length range, the difference in $k$ caused by a strain less than 0.01 is estimated to be less than 10%. Considering the uncertainties involved in experimental measurements (~20%), it is difficult to distinguish the strain effects as strains are often smaller than 0.01. For $L$ > 500 μm, the trend changes, i.e., larger strain gives higher $k$. Strong dependence of $k$ on strain magnitude and sample size results from a competition between boundary and intrinsic three-phonon scattering. Our analysis shows that the mode heat capacities of ZA phonons decrease due to the decrease of phonon number per unit frequency. For a ZA mode $\lambda$, its contribution $k_\lambda^{total}$ to $k$ satisfies $1/k_\lambda^{total} = 1/k_\lambda^{intrinsic} + 1/k_\lambda^{boundary}$, where $1/k_\lambda^{intrinsic}$ and $1/k_\lambda^{boundary}$ are resistivities corresponding to the intrinsic phonon-phonon scattering and boundary scattering, respectively. At a fixed sample length, increasing strain decreases $k_\lambda^{boundary}$ due to the decrease in mode heat capacities of ZA phonons while enhances $k_\lambda^{intrinsic}$ due to the increase in lifetimes of ZA phonons. Therefore, for applicable sample sizes in the range $1 < L \leq 500$ μm for which $k_\lambda^{intrinsic}$ and $k_\lambda^{boundary}$ are comparable, their opposite variation trends will result in a peak enhancement to $k_\lambda^{total}$ by strain, as observed in $k$. For $L > 500$ μm, we find intrinsic three-phonon scattering rates dominate compared with boundary scattering rates for the whole frequency range considered. Therefore, the increase of $k_\lambda^{intrinsic}$ determines the improvement of $k_\lambda^{total}$ and $k$. This result is consistent with those shown in Fig.1(a) for which the boundary scattering is neglected totally. Interestingly, the tensile-strain-induced enhancement of $k$ is in contrast to those reported for other carbon-based materials such as 3D diamond [40] and 1D carbon nanotubes [30,45], wherein tensile strains reduce $k$ through phonon softening [45]. While the softening of LA, TA and optic phonons in graphene is indeed observed here, we find hardening of the ZA modes, i.e., higher ZA frequencies and low-frequency group velocities for the strain levels considered. The ZA phonon hardening coupled with decreasing anharmonic IFCs give increased lifetimes of ZA phonons and enhancement of $k$. We note that, in calculations for strained graphene going from finite size to infinity, using the intrinsic three-phonon scattering rates from finite $q$-sampling density will result in a false convergence for $k$, as shown by cases $\varepsilon=0.0025$ and 0.01 in Fig. 4(b). Theoretically, $N_1 \to \infty$ is required to match $L \to \infty$ and present the divergence for strained graphene. Our tests using larger $q$-sampling density ($N_1>301$) give the uniform results in Fig. 4(b) except that the false convergence plateau for $k$ occurs at larger length, validating our judgments on $k$ variations with size and strain. We note that our investigation is based on three-phonon scattering theory. The high-order phonon scattering (*e.g*, the four-phonon scattering) effect on the *k* convergence of strained graphene is an open question and beyond the range of this work. Considering the high-order phonon scattering strength is usually regarded to be a few orders of magnitude lower [38] that that of three-phonon scattering, the trends presented in Fig.4(b) should not change because the three-phonon scattering strength dominates those trends.

In summary, based on rigorous first-principles lattice dynamics calculations we present a comprehensive picture of phonon thermal transport in unstrained and strained graphene. Good agreement between calculated *k* and experimental data validates the present approach for calculating *k* in 2D systems. We find that the intrinsic room temperature *k* converges for unstrained graphene but diverges for strained graphene with system size. Analysis based on the phonon lifetimes confirms this finding. For unstrained graphene, we conclude that centimeter-order MFP of ZA phonons is responsible for the significant size effect observed in previous experiments. The ZA phonons dominate thermal transport in graphene below 3000 K. For finite strained graphene, tensile strain hardens the flexural modes, increases their lifetimes and causes unusual dependences of thermal conductivity on sample size and strain due to the competition between the boundary and phonon-phonon scattering; Once the sample larger than 500 μm, increasing strain or size may effectively enhances its *k*.

**Acknowledgments**

We are thankful for the financial support from the Hong Kong General Research Fund under Grant Nos. 623212 and 613413. L. L. acknowledges support from the U. S. Department of Energy, Office of Science, Office of Basic Energy Sciences,